\pdfoutput=1

\documentclass[11pt]{article}

\usepackage[final]{acl}
\usepackage{times}
\usepackage{booktabs}
\usepackage{latexsym}
\usepackage{subcaption}
\usepackage{multirow}
\usepackage[ruled,vlined]{algorithm2e}
\usepackage{placeins}
\SetKwInOut{Input}{Input}
\SetKwInOut{Output}{Output}
\usepackage{amsmath}
\usepackage{float}

\usepackage[T1]{fontenc}

\usepackage[utf8]{inputenc}

\usepackage{microtype}

\usepackage{inconsolata}
\usepackage{booktabs}
\usepackage[table]{xcolor} 
\usepackage{graphicx}

\title{DiTReducio: A Training-Free Acceleration for DiT-Based TTS via Progressive Calibration}

\author{
  Yanru Huo$^2$, Ziyue Jiang$^1$, Zuoli Tang$^3$, Qingyang Hong$^2$, Zhou Zhao$^1$\thanks{Corresponding author} \\
  \vspace{0.5em}
  $^1$Zhejiang University, $^2$Xiamen University, $^3$Wuhan University\\
  \texttt{hyrrrr0661@xmu.edu.cn}
}

\begin{document}
\maketitle
\begin{abstract}
While Diffusion Transformers (DiT) have advanced non-autoregressive (NAR) speech synthesis, their high computational demands remain an limitation. Existing DiT-based text-to-speech (TTS) model acceleration approaches mainly focus on reducing sampling steps through distillation techniques, yet they remain constrained by training costs. We introduce DiTReducio, a training-free acceleration framework that compresses computations in DiT-based TTS models via progressive calibration. We propose two compression methods, \textbf{Temporal Skipping} and \textbf{Branch Skipping}, to eliminate redundant computations during inference. Moreover, based on two characteristic attention patterns identified within DiT layers, we devise a pattern-guided strategy to selectively apply the compression methods. Our method allows flexible modulation between generation quality and computational efficiency through adjustable compression thresholds. Experimental evaluations conducted on F5-TTS and MegaTTS 3 demonstrate that DiTReducio achieves a 75.4\% reduction in FLOPs and improves the Real-Time Factor (RTF) by 37.1\%, while preserving generation quality. 

\end{abstract}

\section{Introduction}

Recent advances in TTS synthesis have enabled the generation of highly realistic and natural speech, with applications including virtual assistants, audiobooks, and digital avatars. The autoregressive (AR) models~\cite{wang2023neural,xin2024rall,chen2024vall,anastassiou2024seed,du2024cosyvoice,song2025ella,huang2023make,wang2025spark,deng2025indextts} and NAR TTS~\cite{wang2024maskgct,huang2022generspeech} have demonstrated robust zero-shot capabilities, especially those based on DiT~\cite{mehta2024matcha,ju2024naturalspeech,chen2024f5,eskimez2024e2,jiang2025megatts} achieve accelerated inference while maintaining audio generation quality through high-performance computational parallelization. These benefits have facilitated their widespread adoption in real-world applications.

Despite their benefits, DiT-based models fundamentally face architectural limitations. While effectively capturing long-range dependencies, the Transformer's self-attention mechanism results in quadratic time and space complexity. This computational burden is further exacerbated by the multi-step denoising process and Classifier-Free Guidance (CFG) techniques \cite{ho2022classifier} employed in these models. While some lightweight DiT-based TTS systems have achieved efficient inference, their computational requirements still exceed the limits of on-device deployment and real-time interactive applications, highlighting the need for inference acceleration methods.

Prior works address these computational challenges through three main approaches: (1) optimizing diffusion sampling using advanced samplers or distillation to reduce inference steps~\cite{lu2022dpm,lu2022dpm++,salimans2022progressive}; (2) applying model compression techniques such as quantization and pruning ~\cite{shang2023post,wan2025pruning}; and (3) implementing attention mechanism optimizations, such as sparse attention~\cite{zaheer2020big,hassani2023neighborhood} and token-wise methods~\cite{bolya2023token,saghatchian2025cached} that may face compatibility issues with FlashAttention~\cite{dao2022flashattention}. While numerous innovations in efficient inference~\cite{yuan2024ditfastattn} have emerged in image and video generation, comparable advancements in TTS remain limited, highlighting a critical research gap.

\begin{figure*}[t]
    \centering
    \includegraphics[width=\textwidth]{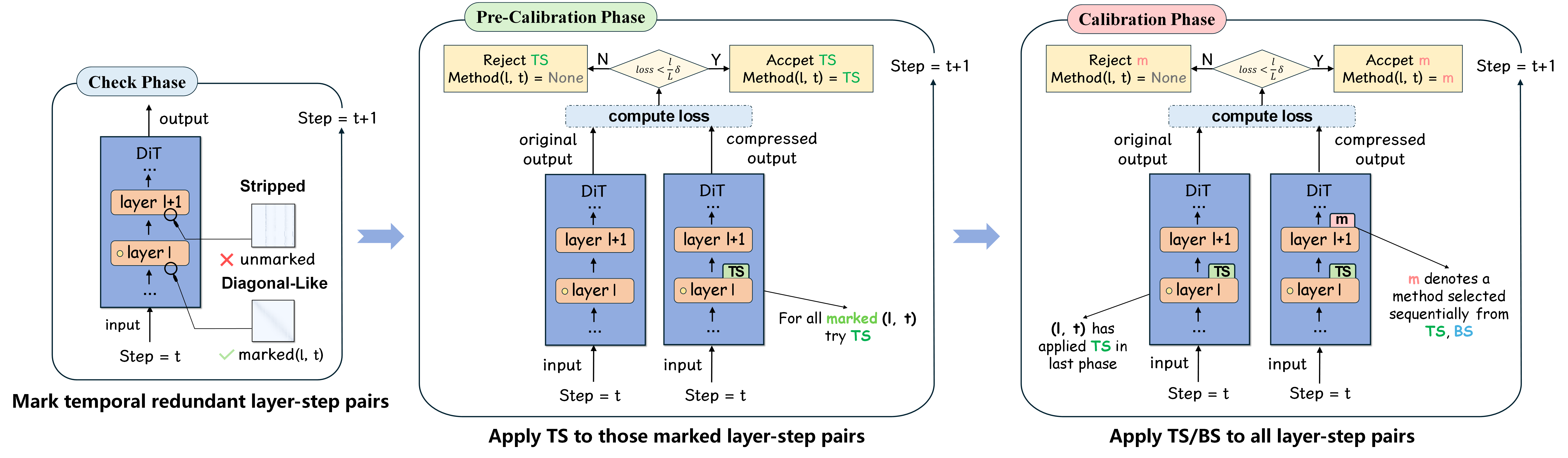} 
    \caption{\textbf{Overview of DiTReducio}. In the Check Phase, we identify a subset of highly temporally redundant layer-step pairs by detecting diagonal-like attention patterns. In the Pre-Calibration Phase, we apply TS to those identified pairs and retain only those for which the resulting output loss remains below a dynamical threshold. Finally, in the Calibration Phase, both TS and BS are applied across all layer-step pairs under the same loss constraint. This procedure yields a model‑specific inference acceleration strategy.}
    \label{fig:method_overview}
\end{figure*}

Our goal is to develop a training-free approach for quality-controllable acceleration in DiT-based TTS models. Inspired by prior observations of computational redundancies in the DiT architectures~\cite{yuan2024ditfastattn}, we systematically investigate two key phenomena. 

Firstly, through detailed analysis of DiT architectures, we identify two notable forms of computational redundancy manifested during model inference at specific layer-timestep combinations (referred to as layer-step pairs below). The first, temporal redundancy, is characterized by high output similarity across adjacent diffusion denoising timesteps in attention mechanisms and feed-forward networks (FFNs) at particular layers. The second, branch redundancy, emerges as similar outputs between conditional and unconditional branches at particular layer-step pairs. To leverage these redundancies, we introduce two tailored strategies: \textbf{Temporal Skipping (TS)} and \textbf{Branch Skipping (BS)}. TS exploits temporal redundancy through caching and reusing of computational results across timesteps, while BS derives the unconditional branch output using the computed conditional branch output and cached branch residuals from the previous timestep.

To further investigate the underlying mechanisms of these redundancies, we analyze attention heatmaps in specific layer-step pairs and uncover two distinct patterns, diagonal-like patterns and striped patterns. The diagonal-like patterns are characterized by tokens primarily attending to their neighboring tokens. This suggests that these layers focus on local acoustic refinement, such as prosody and spectral details within short speech segments at these timesteps. While the interpretability of striped patterns remains challenging, our empirical studies demonstrate their crucial role in maintaining the overall speech generation quality, particularly in preserving the coherence and naturalness of the synthesized speech.

Building on these insights, in this paper, we propose DiTReducio, a systematic progressive calibration framework for efficient DiT-based TTS inference. As Figure~\ref{fig:method_overview} shows, the framework operates through the following sequential phases:

\begin{enumerate}
    \item \textbf{Check Phase}: Identifying layer-step pairs exhibiting diagonal-like attention patterns as highly temporally redundant.
    \item \textbf{Pre-Calibration phase}: Applying the TS strategy selectively to the marked layer-step pairs.
    \item \textbf{Calibration phase}: Building upon the pre-calibrated model, applying both TS and BS strategies across all layer-step pairs while preserving generation quality.
\end{enumerate}

Experimental results demonstrate that our approach can achieve a 1.6× improvement in RTF while maintaining controllable generation quality. As a training-free and plug-and-play solution, DiTReducio can seamlessly integrate with existing acceleration methods. Also, the framework's adaptability in balancing acceleration and quality preservation makes it particularly advantageous for large-scale TTS deployments.


\section{Related Work}
\subsection{Diffusion-based Speech Synthesis}
The emergence of diffusion models has challenged the long-standing dominance of AR models in speech synthesis. Leveraging NAR generation paradigms, diffusion models enhance generation efficiency while preserving high synthesis quality. Early efforts such as Diff-TTS~\cite{jeong2021diff} pioneered the application of diffusion models to speech synthesis, demonstrating their feasibility. Guided-TTS~\cite{kim2022guided} further introduced CFG mechanisms, substantially improving the controllability and naturalness of generated speech.

With the development of Latent Diffusion Models (LDM)~\cite{rombach2022high}, especially the emergence of DiT~\cite{peebles2023scalable}, diffusion-based speech synthesis has entered a new and promising phase. These models~\cite{lee2024ditto,eskimez2024e2,chen2024f5,du2024cosyvoice,jiang2025megatts} fully exploit the structural parallelism of Transformer architectures within the latent diffusion framework, achieving both efficient training and inference as well as robust zero-shot speech generation. This dual advantage of computational efficiency and reliable generation has made them well-suited for integration with multi-modal large language models~\cite{xu2025qwen2} in practical applications.

\subsection{Acceleration of Diffusion Model}
Despite the excellent performance of diffusion models in generation tasks, their inherent multi-step denoising nature leads to high computational costs, constraining the practical application of end-to-end speech synthesis. Current mainstream acceleration methods primarily include model distillation, sampler optimization, quantization, and pruning. Among these, model distillation~\cite{salimans2022progressive,sauer2024adversarial} techniques transfer the capabilities of complex teacher models to lightweight student models, reducing sampling steps. However, such methods require additional training overhead and depend on the performance of the teacher model. Improvements to samplers~\cite{lu2022dpm,lu2022dpm++} optimize noise schedules and achieve generation quality comparable to that of thousand-step sampling with significantly fewer sampling steps. These methods have the advantage of being training-free and are relatively mature. While quantization~\cite{li2023q,he2023ptqd} and pruning~\cite{castells2024ld} can improve inference efficiency, they lead to unpredictable generation quality degradation, and pruning methods also require additional training efforts.

Some works in image and video generation focus on optimizing the attention mechanism and propose token-wise acceleration algorithms~\cite{bolya2022token,kong2022spvit,xing2024pyramiddrop}. However, these methods require training a token selector or calculating attention heatmaps at each inference step, which causes compatibility issues with efficient computation libraries such as FlashAttention, and their high implementation complexity limits practical applications.

In speech synthesis, inference acceleration is primarily achieved through distillation~\cite{huang2022prodiff,li2024dmdspeech,guan2024reflow}. Apart from distillation, architectural improvements offer an alternative approach. DiffGAN-TTS~\cite{liu2022diffgan} introduces Generative Adversarial Networks (GANs) to simulate denoising distributions, enabling faster inference. While such methods can significantly improve speed, they may introduce additional training complexity.

Despite the progress, training-free and plug-and-play acceleration solutions remain limited in diffusion-based speech synthesis. Achieving low-cost acceleration while preserving generation quality remains an urgent challenge in diffusion-based speech synthesis. 

\begin{figure*}[t]
    \centering
    \includegraphics[width=0.9\textwidth]{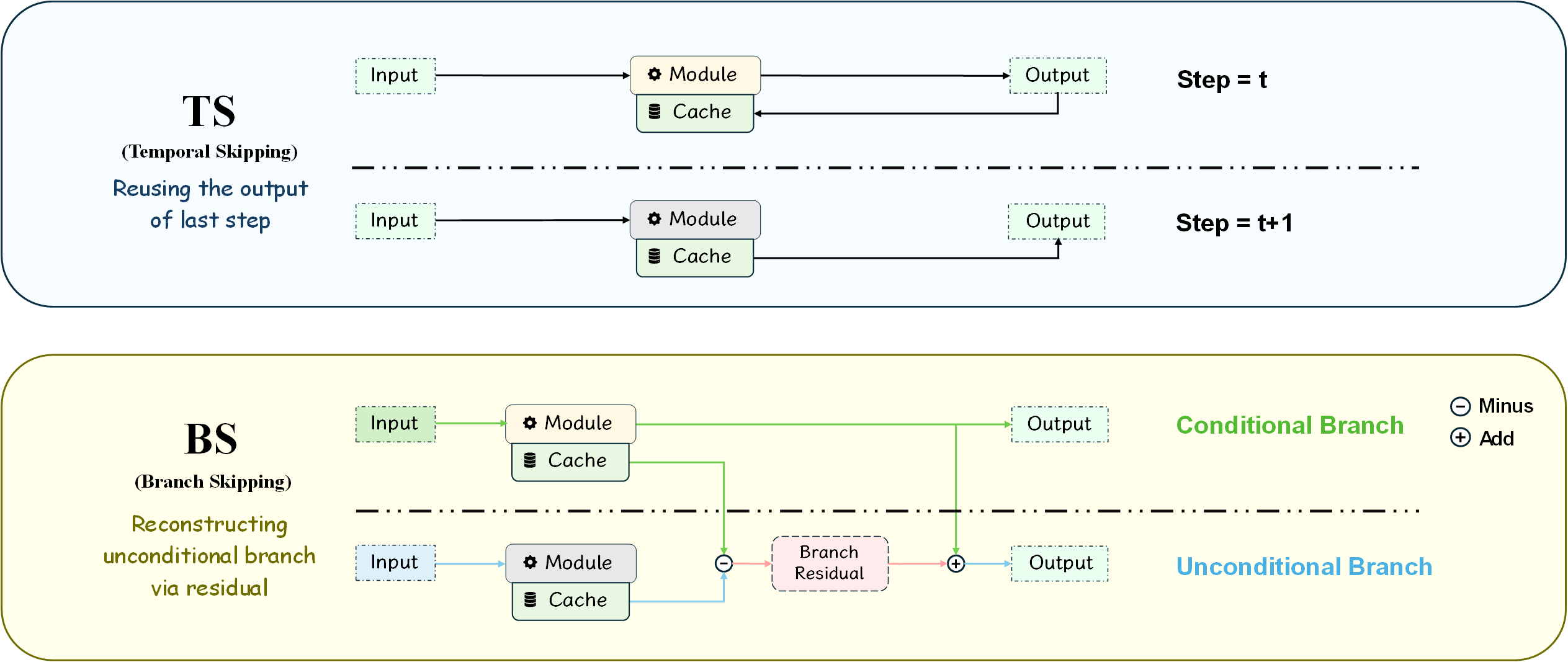} 
    \caption{\textbf{Comparison of the workflows of TS, BS.} The cache is updated only when TS is not applied by the corresponding module.}
    \label{fig:method_methods}
\end{figure*}

\section{Method}

\subsection{Overview}\label{sec:method_overview}
In this section, we present the methodology of DiTReducio. In Section~\ref{sec:method_compression-methods}, we systematically investigate two types of redundancy, temporal redundancy and branch redundancy, arising during model inference, and introduce corresponding compression methods. In Section~\ref{sec:method_attention-pattern}, we analyze the self-attention patterns within DiT layers and demonstrate their correlation with temporal redundancy across layer-step pairs. In Section~\ref{sec:method_DiTReducio}, we propose DiTReducio, a progressive calibration framework that iteratively explores compression opportunities through three inference passes. The framework records effective acceleration strategies based on observed redundancy, enabling plug-and-play deployment in DiT-based speech synthesis.

\subsection{Compression Methods}\label{sec:method_compression-methods}
Redundant computations during diffusion model inference pose a major bottleneck to inference speed. Previous studies~\cite{ma2024learning,zhao2024real}~have shown that diffusion models for image and video generation exhibit substantial temporal redundancy across multiple timesteps. In our work, we demonstrate that diffusion models for speech synthesis similarly exhibit extensive temporal redundancy and, under CFG, also display measurable branch redundancy.

\paragraph{Temporal redundancy}Temporal redundancy refers to a high similarity between the outputs of a given module at adjacent timesteps during model inference. Figure~\ref{fig:tr-of-models} in Appendix~\ref{sec:app_red} indicates the cosine similarity heatmaps of the attention and feed-forward modules at different timesteps for both F5-TTS and MegaTTS 3. This observation reveals three key insights: (1) outputs across different timesteps exhibit strong similarity; (2) output similarity increases as timesteps become temporally closer, especially for adjacent ones; (3) temporal redundancy exists across multiple modules.

Based on this, we introduce the \textbf{Temporal Skipping} (TS) strategy. As illustrated in Figure~\ref{fig:method_methods}, TS caches an output from a specific module at the preceding timestep and reuses it in subsequent steps to avoid temporally redundant computation. Formally, let $O_t$ denote as the output of a module at timestep $t$. Under TS, we have:
\begin{equation}
O_t = O_{t-1}.
\end{equation}

\paragraph{Branch redundancy} Branch redundancy arises in CFG-based diffusion models when the conditional branch output $O_t^c$ and the unconditional branch output $O_t^u$ of a module at a given timestep are highly similar. Figure~\ref{fig:br-of-models} in Appendix~\ref{sec:app_red} indicates the cosine similarity heatmaps between $O_t^c$ and $O_t^u$ for the attention and feed-forward modules in both F5-TTS and MegaTTS 3, confirming the presence of obvious branch redundancy. This observation reveals the following insights: (1) outputs across branches have high similarity but in particular steps; (2) branch redundancy exists across multiple modules.

To address this, we propose the \textbf{Branch Skipping} (BS) strategy. Unlike the direct reuse mechanism in the TS strategy, branch redundancy is comparatively less obvious than temporal redundancy; therefore, BS exploits both types of redundancy by computing the branch residual. As shown in Figure~\ref{fig:method_methods}, under BS, only the conditional branch is executed, and the unconditional branch output for the current timestep is reconstructed using the conditional branch output and the branch residual. Here, the branch residual is computed as the difference between the cached unconditional and conditional branch outputs. Let the output of a module at timestep $t$ be $O_t = \mathrm{Concat}\bigl(O_t^c, O_t^u\bigr)$, under BS, the output becomes
\begin{equation}
O_t = \mathrm{Concat}\bigl(O_t^c, O_{t-1}^u-O_{t-1}^c + O_{t}^c\bigr),
\end{equation}

This formulation skips the redundant branch computation while ensuring that the resulting output closely approximates the original.


\subsection{Attention Pattern}\label{sec:method_attention-pattern}

Early identification and application of TS on temporally redundant layer-step pairs avoids forcing them into suboptimal compression strategies during the subsequent greedy-based calibration phase. Therefore, we focus on developing accurate methods for detecting such redundancy. In fact, the degree of temporal redundancy depends not only on the interval between timesteps but also on the distinct functional roles of internal layers in generation tasks. DeepCache~\cite{ma2024deepcache} mentioned that shallow layers in diffusion models construct the overall outline, whereas deeper layers are responsible for synthesizing fine details. In our work, we further explore the relationship between the attention patterns of each layer in the DiT and the distinct functional roles of those layers in the inference process, leveraging this connection to efficiently identify highly temporally redundant layer-step pairs.

\begin{figure}[h]
    \begin{subfigure}[t]{0.48\linewidth}
      \includegraphics[width=\columnwidth]{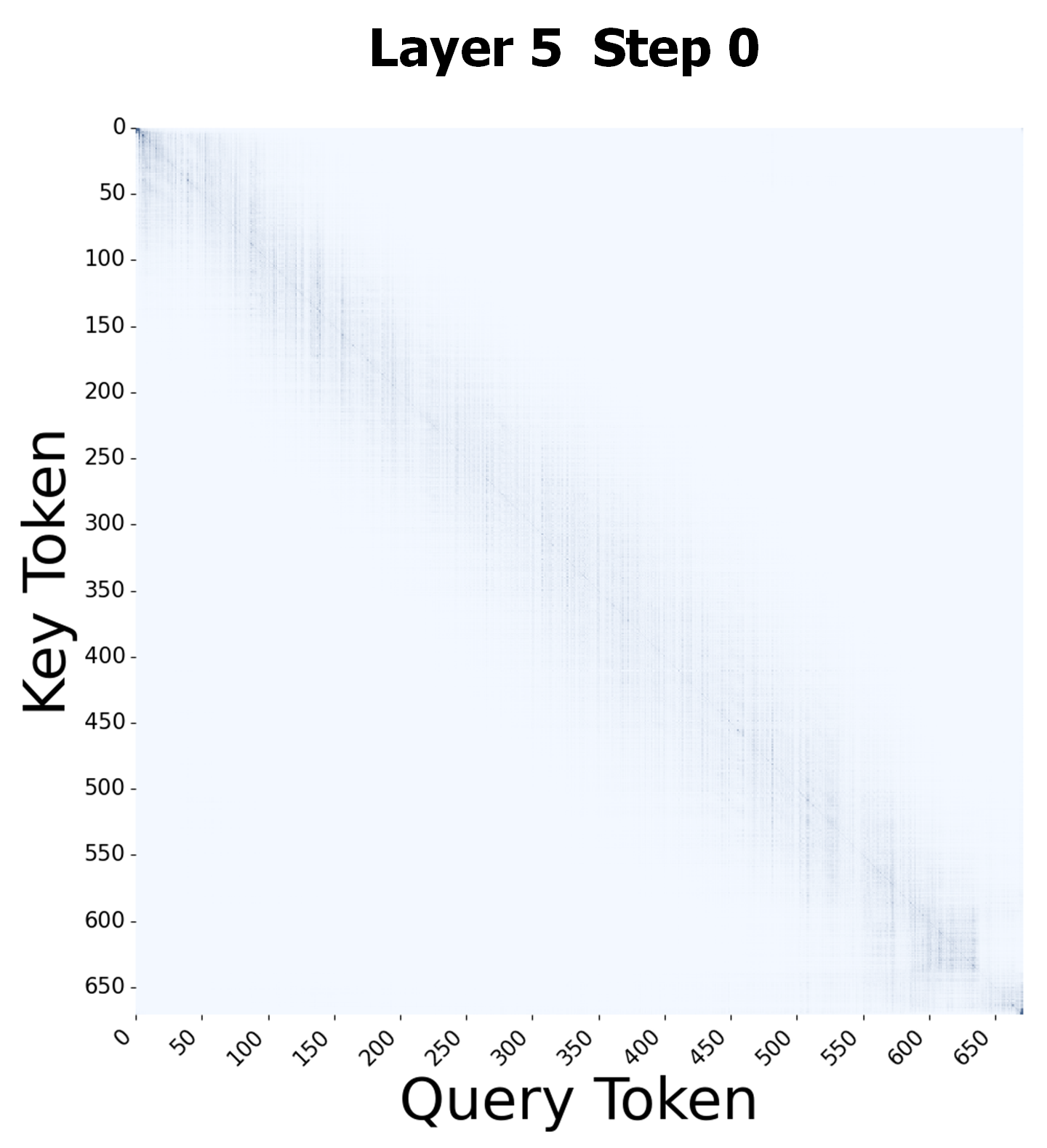}
      \caption{Diagonal-like Pattern}
    \end{subfigure}
    \hfill
    \begin{subfigure}[t]{0.48\linewidth}
      \includegraphics[width=\columnwidth]{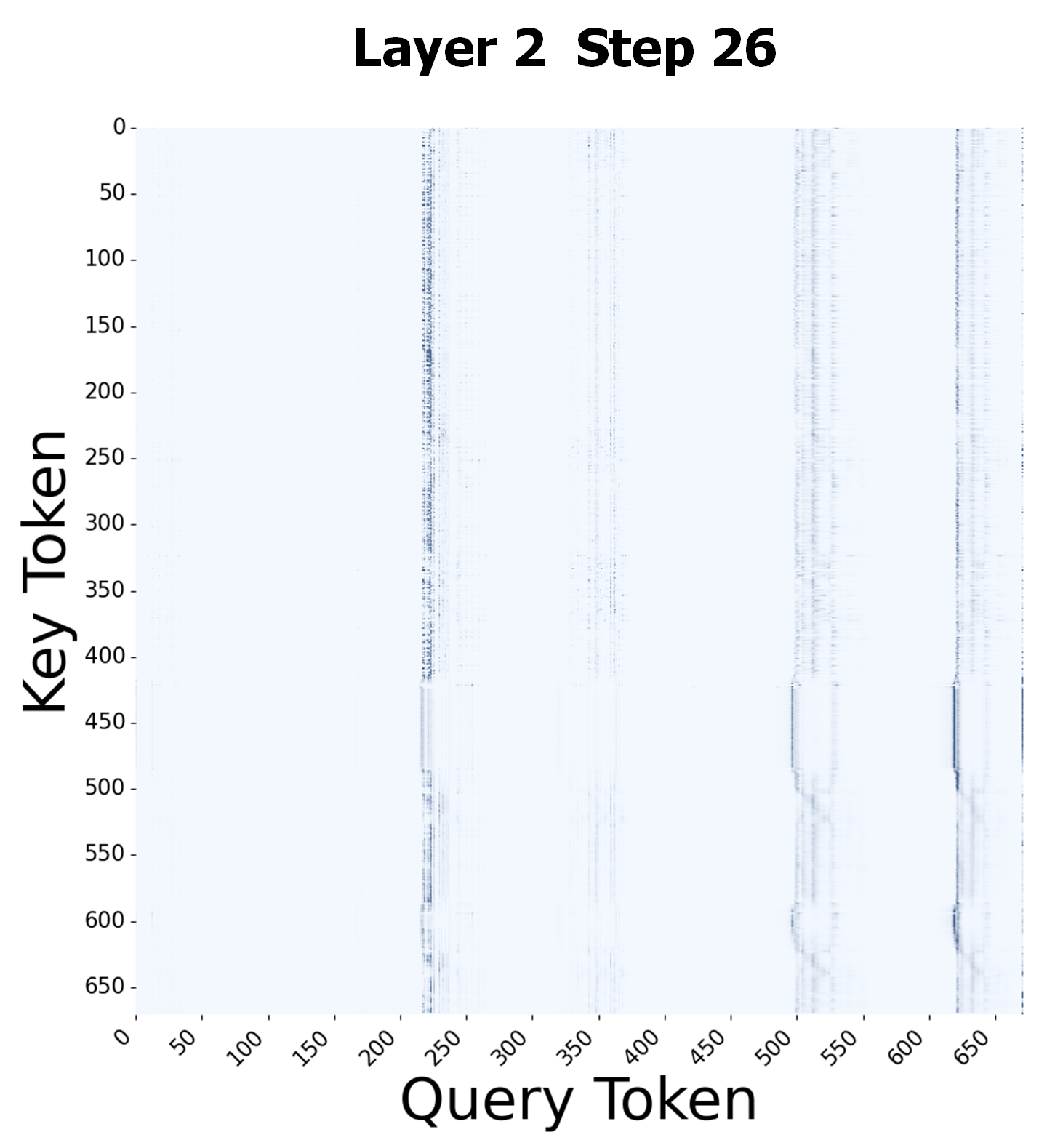}
      \caption{Striped Pattern}
    \end{subfigure}
  \caption {\textbf{Attention Patterns in F5-TTS inference}: (a) Diagonal-like patterns in both conditional and unconditional branches. (b) Striped patterns in both branches.}
  \label{fig:intro_patterns}
\end{figure}

The attention heatmaps in the DiT layers exhibit two distinct patterns, diagonal-like and striped. These patterns indicate potential functional distinctions among layer-step pairs. Figure~\ref{fig:intro_patterns} presents these patterns during inference in the F5-TTS model: (a) shows the diagonal-like pattern for layer 5 at timestep 0, while (b) displays the striped pattern for layer 2 at timestep 26. In the context of speech synthesis, we interpret the diagonal-like pattern as handling fine-grained local details, thereby enhancing speech fidelity and clarity. Although the exact function of the striped pattern requires further investigation, our empirical analysis suggests that striped patterns are crucial and less redundant.

\begin{figure}[h]
    \centering
    \includegraphics[width=\columnwidth]{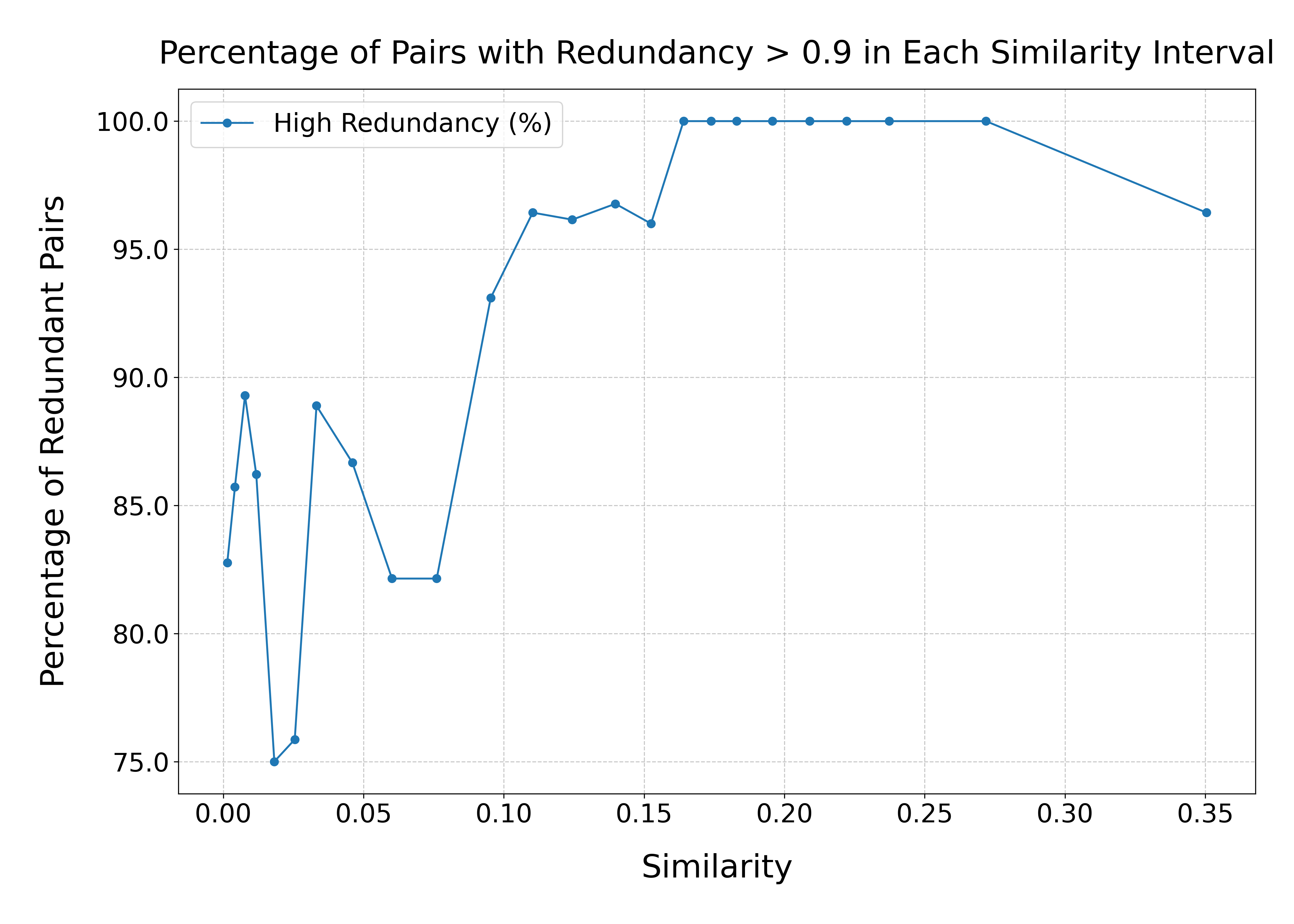} 
    \caption{\textbf{Percentage of temporally redundant layer-step pairs in each similarity interval in F5-TTS}.}
    \label{fig:redundant-ratio-analyse}
\end{figure}

To further analyze the diagonal-like attention pattern, we collected the cosine similarity between attention heatmaps and diagonal matrices, as well as the corresponding temporal redundancy across all layer-step pairs in F5-TTS. The greater the similarity, the more closely the layer-step pair aligns with a diagonal-like pattern. For layer $l$ at timestep $t$, let $O$ denote the original output of the model and $O_{l,t}$ represent the final model output when layer $l$ uses the cached output from the timestep $t-1$ instead of recomputing it at the current timestep $t$. Based on these outputs, we define the temporal redundancy as:
\begin{equation}
\text{\small
$\displaystyle
R_{l,t} = 1 - \frac{1}{b\cdot n\cdot d}\sum\limits_{k=1}^b \sum\limits_{i=1}^n \sum\limits_{j=1}^d ||O_{k,i,j} - O'_{k,i,j}||_1
$%
},
\end{equation} where $b$ denotes the batch size, $n$ the sequence length, and $d$ the feature dimension. This metric measures the mean absolute error between $O$ and $O_{l,t}$. Layer-step pairs with redundancy above 0.9 are marked as highly temporally redundant. Figure~\ref{fig:redundant-ratio-analyse} illustrates the percentage of temporally redundant layer-step pairs in each similarity interval in F5-TTS. At lower similarity, specifically below 0.1, the percentage of redundant pairs remains relatively low, fluctuating between 75\% and 90\%. As the similarity increases from 0.1 to 0.35, the redundancy percentage exhibits a notable upward trend, reaching a peak of nearly 100\% between 0.15 and 0.35. The phenomenon suggests a tendency that: (1) DiT layers exhibiting the diagonal-like patterns are more likely to demonstrate higher temporal redundancy during inference; (2) conversely, layers with striped patterns tend to exhibit lower temporal redundancy. 

\begin{algorithm*}[htbp]
    \caption{Calibration Phase of DiTReducio}
    \SetKwInOut{Input}{Input}
    \SetKwInOut{Output}{Output}
    
    \Input{Model $M$, Calibration threshold $\delta$, total number of layers $L$}
    \Output{Strategy table $\text{strategy\_table}[T][L]$}

    Initialize $\text{strategy\_table}$ as a $T \times L$ matrix filled with $\text{NONE}$\;
    
    During the sampling process of the model, at each timestep $t$: 

    \For{layer $l \leftarrow 1$ \KwTo $L$}{
        
        \For{method $m \in \{\text{TS}, \text{BS}\}$}{
            $O \gets$ model output at timestep $t$ without compression at layer $l$\;
            
            $O' \gets$ model output at timestep $t$ with method $m$ applied to layer $l$\;
            
            $\epsilon \gets \frac{1}{n} \sum |O - O'|$, where $n$ is the number of elements in the tensor\;
            
            \If{$\epsilon < \frac{l}{L} \cdot \delta$}{
                Update $\text{strategy\_table}[t][l] \gets m$\;
            }
        }
    }
\label{alg:calibration}
\end{algorithm*}

\subsection{DiTReducio}\label{sec:method_DiTReducio}

Previous studies~\cite{sun2024unveiling} analyzed the computational redundancy of prevalent diffusion transformers relative to their inputs, revealing that internal redundancy patterns are input-agnostic—i.e., they depend primarily on model architecture and model parameters rather than specific input content. This property enables the construction of model-agnostic acceleration strategies via a limited number of inference runs. Accordingly, we introduce DiTReducio, which achieves remarkable acceleration by performing only three iterative inference passes and trade off generation quality and inference speed with a predefined compression threshold. DiTReducio comprises three phases: \textbf{Check Phase}, \textbf{Pre-Calibration Phase}, and \textbf{Calibration Phase}.  

\paragraph{Check Phase} The goal of this phase is to mark highly temporally redundant layer-step pairs before calibration. Leveraging the correlation between attention patterns and redundancy revealed in Section~\ref{sec:method_attention-pattern}, we compute the cosine similarity $s_{l,t}$ between each layer’s attention heatmap and the identity matrix before the forward of layer $l$ at step $t$. After inference, all $s_{l,t}$ values are sorted in descending order, and the top $q\%$ highest-similarity layer-step pairs are marked.

\paragraph{Pre-Calibration Phase} This phase performs a preliminary calibration based on the Check Phase results. Its procedure mirrors that of the Calibration Phase, except that (1) only the marked layer-step pairs are considered, and (2) only the TS is applied, without the BS. The ablation studies in Section~\ref{sec:ablation} verify that this phase substantially improves the quality of the resulting acceleration strategy.

\begin{table*}[h]
  \setlength\tabcolsep{2.2mm}
  \renewcommand{\arraystretch}{1.22}
  \centering
  \small
  \begin{tabular}{ccccccccc}
    \toprule
    \textbf{Model} & \textbf{Metric} & \multicolumn{7}{c}{\textbf{Threshold}} \\
    \cmidrule(lr){3-9}
    & & T0 & T1 & T2 & T3 & \textbf{T4} & T5 & T6 \\
    \midrule
    \multirow{4}{*}{\textbf{F5-TTS}}
      & SIM-o         & 0.640 & 0.640 & 0.637 & 0.629 & \textbf{0.618} & 0.610 & 0.590 \\
      & WER (\%)      & 2.636 & 2.655 & 2.564 & 2.643 & \textbf{2.634} & 2.661 & 2.900 \\
      & RTF           & 0.178 & 0.165 & 0.149 & 0.138 & \textbf{0.129} & 0.120 & 0.112 \\
      & Ops Ratio (\%)& 100.00 & 82.59 & 66.38 & 55.09 & \textbf{45.58} & 39.26 & 34.42 \\
    \midrule
    \multirow{4}{*}{\textbf{MegaTTS 3}}
      & SIM-o         & 0.750 & 0.750 & 0.748 & 0.743 & \textbf{0.734} & 0.691 & 0.626 \\
      & WER (\%)      & 3.112 & 3.112 & 3.110 & 3.073 & \textbf{3.095} & 3.133 & 3.030 \\
      & RTF           & 0.396 & 0.395 & 0.359 & 0.287 & \textbf{0.224} & 0.176 & 0.156 \\
      & Ops Ratio (\%)& 100.00 & 98.87 & 88.02 & 68.19 & \textbf{48.94} & 33.88 & 27.52 \\
    \bottomrule
  \end{tabular}
  \caption{\textbf{Performance comparison between F5-TTS and MegaTTS 3 under varying compression thresholds.} The bold column (T4) represents the optimal threshold, at which the models achieve substantial acceleration while maintaining generation quality within acceptable bounds. \emph{Ops Ratio} denotes the ratio of FLOPs after compression relative to the baseline (uncompressed) model, indicating the extent of computational reduction. The data is evaluated on a single Nvidia 3090 GPU.}
  \label{tab:core-results}
\end{table*}

\paragraph{Calibration Phase} This phase performs a comprehensive calibration following the Pre-Calibration Phase. Since the impact of compressing deep layers is greater than that of compressing shallow layers, under a given global calibration threshold $\delta$, we introduce a dynamic threshold that controls the computational compression for the layer-step pair $(l,t)$ using the threshold $\frac{l}{L}\delta$, where $L$ is the total number of layers in the DiT. The algorithm is outlined in Algorithm~\ref{alg:calibration}. During the model's inference at step $t$, for layer $l$, we first compute the entire model output $O$ without applying any acceleration strategy to pair $(l,t)$. Then we sequentially apply both TS and BS strategies to the attention and feed-forward modules of the layer to obtain the compressed output $O'$, prioritizing TS due to its higher computational savings. Next, we compute the mean absolute error between $O$ and $O'$. If this value is less than the threshold $\frac{l}{L}\delta$, we record the applied strategy used for $(l,t)$; otherwise, we record that no acceleration strategy is used for $(l,t)$. Notably, if $(l,t)$ has been selected for TS application during the Pre-Calibration Phase, we skip the calibration for that pair.

\section{Experiments}

\subsection{Settings}

\paragraph{Model} We evaluate DiTReducio on F5-TTS~\cite{chen2024f5} and MegaTTS 3~\cite{jiang2025megatts}, both of which employ DiT for conditional flow matching. To implement our compression method, we make several modifications to the model. Implementation details are provided in the Appendix~\ref{subsec:implementation}.

\paragraph{Dataset \& Task} We utilize the LibriSpeech-PC-test-clean subset~\cite{meister2023librispeech} comprising 1,127 samples, following F5-TTS. We assess the performance of the model under DiTReducio within the cross-sentence task paradigm, where models generate speech with consistent speaker characteristics based on a reference text, a speaker prompt, and a corresponding transcription.

\paragraph{Metric} We adopt the evaluation metrics from F5-TTS to assess generation quality. The speaker similarity-objective (SIM-o) is computed using the WavLM-large-based model~\cite{chen2022large} to compute cosine similarity between features extracted from synthesized and reference audio. The Word Error Rate (WER) is calculated by comparing the reference text against transcriptions generated by Whisper-large-v3~\cite{radford2023robust}. To evaluate acceleration performance, we adopt the RTF to quantify inference speed, and use the total floating-point operations (FLOPs) as an indicator of computational costs. For F5-TTS, the RTF is measured based on the model's total inference time. For MegaTTS 3, only the DiT inference time is considered, as DiT is not the sole bottleneck in the overall inference pipeline. 

\begin{figure}[h]
    \centering
    \includegraphics[width=\columnwidth]{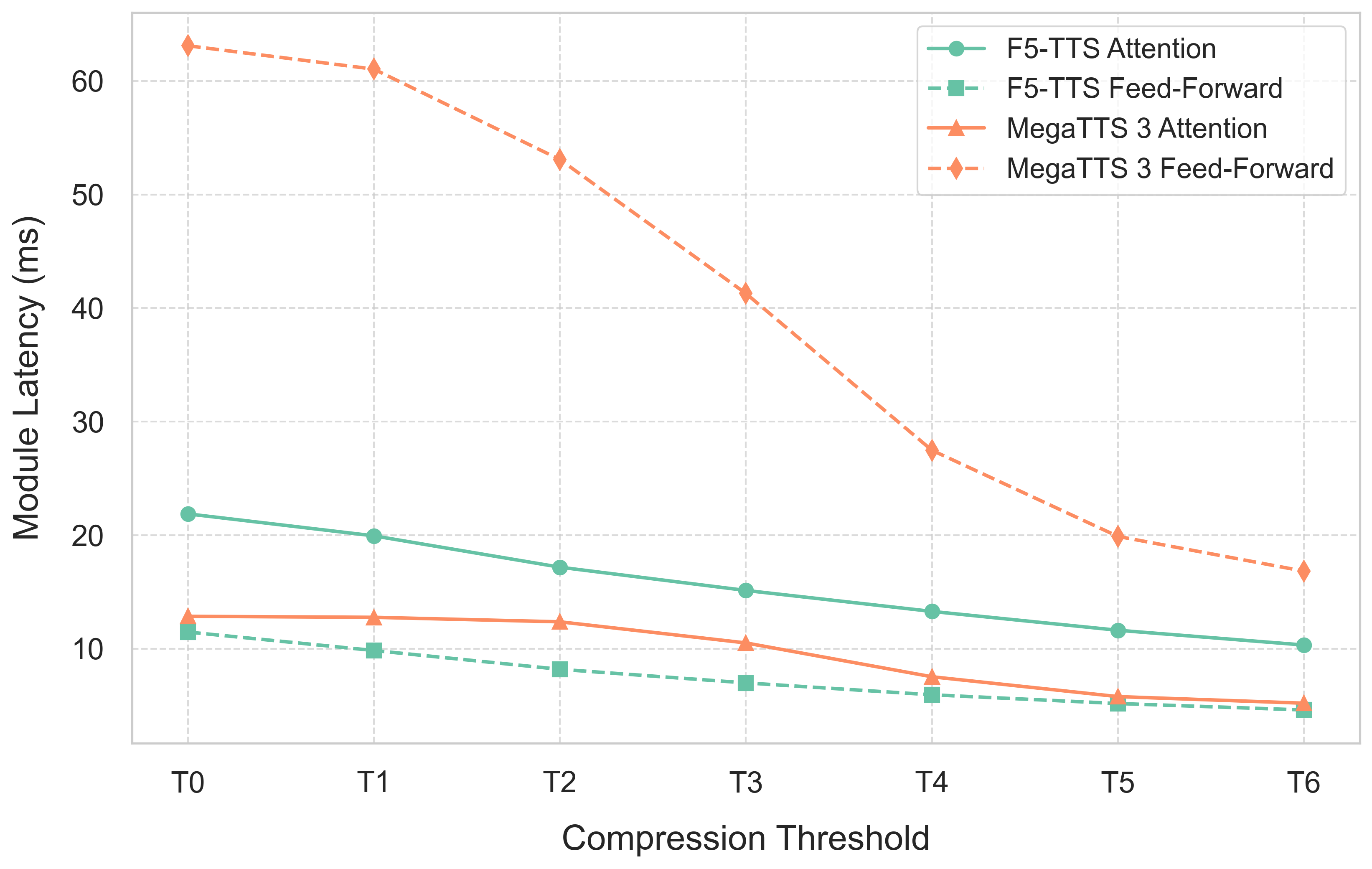} 
    \caption{\textbf{Attention and feed-forward module latencies of F5-TTS and MegaTTS 3}.}
    \label{fig:module_latency}
\end{figure}

\paragraph{Evaluation} All evaluations are conducted across 5 random seeds (42, 3407, 666, 3954, 3962), with results averaged. For both F5-TTS and MegaTTS 3, we evaluate 6 distinct compression thresholds (denoted as T1 through T6 in ascending order, where T0 represents the uncompressed baseline). The thresholds are uniformly distributed, with maximum values of 0.3 and 1.2 for F5-TTS and MegaTTS 3, and the maximum threshold resulting in approximately 10\% degradation in SIM-o. In the Check Phase, we identify the top 10\% of layer-step pairs as highly temporally redundant pairs.

\begin{figure*}[t]
    \centering
    \includegraphics[width=0.9\textwidth]{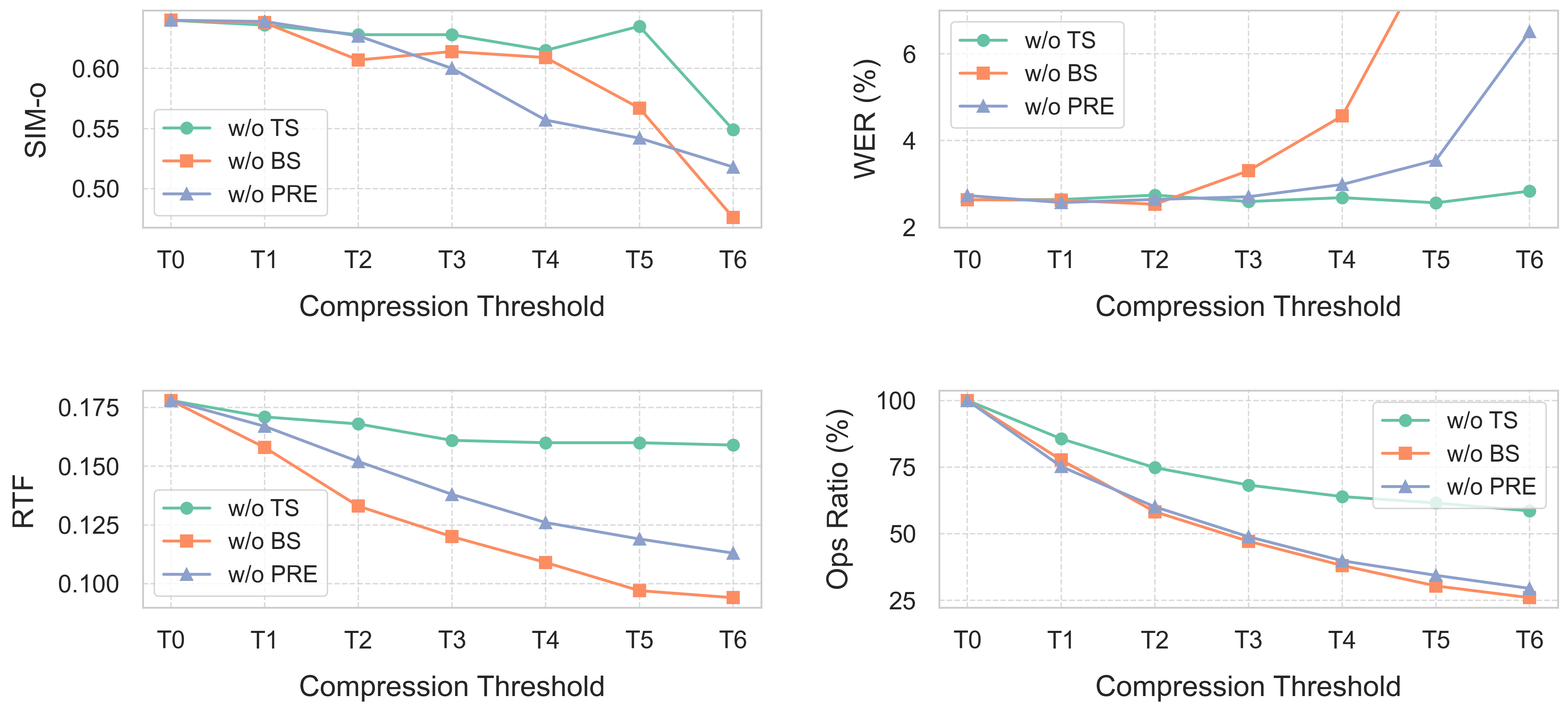} 
     \caption{
       \textbf{Ablation study on F5-TTS.} \textbf{PRE} represents the Check Phase and the Pre-Calibration Phase.
     }
    \label{fig:ablation}
\end{figure*}

\subsection{Results}
\textbf{DiTReducio demonstrates controlled acceleration while preserving the generation quality.} When applying the two lowest thresholds to F5-TTS, the model maintains generation quality comparable to the baseline while achieving significant RTF reduction. At threshold T2, RTF decreases by 16.5\%. With increasing thresholds, the generation quality degradation remains moderate while inference speed improves further, with a maximum RTF reduction of 37.0\%. The similar trend is observed for MegaTTS 3: obvious quality degradation appears only at the highest threshold, while inference efficiency improves more notably.

\textbf{DiTReducio significantly reduce the latency of modules.} We measured the internal attention and feed-forward module latencies of F5-TTS and MegaTTS 3 under varying compression thresholds. As shown in Figure~\ref{fig:module_latency}, both F5-TTS and MegaTTS 3 exhibit considerable decreases in module latency as the compression thresholds increase. At the maximum compression threshold, F5-TTS achieves latency reductions of 52.8\% and 59.8\% for the atention and feed-forward modules, respectively. For MegaTTS 3, the latency reductions reach 60.3\% for the attention module and 63.6\% for the feed-forward module.

\textbf{The selection of compression thresholds is critical.} Our analysis reveals an interesting threshold-dependent behavior: from lower to moderate thresholds (T2 to T4), increasing the threshold yields considerable speedup with minimal quality degradation. However, at higher threshold levels (T5 to T6), the marginal acceleration benefits diminish while quality degradation becomes more evident. This phenomenon suggests that compression approaches its theoretical limit around T4, beyond which further threshold increases may incorrectly identify essential computations as redundant. With appropriate threshold selection, DiTReducio effectively balances inference acceleration and generation quality. Furthermore, different models exhibit varying sensitivities to threshold settings: the F5-TTS shows quality degradation at a compression threshold of 0.3, while MegaTTS 3 retains quality up to a threshold of 0.4, where it achieves notable compression gains.

\subsection{Ablation Study}\label{sec:ablation}
In this section, we investigate the impact of Check Phase, Pre-calibration Phase and two compression methods on DiTReducio's performance.

\textbf{TS strategy is crucial for achieving effective acceleration.} As shown in Figure~\ref{fig:ablation}, ablation results demonstrate that applying only the Branch Skipping (BS) strategy to F5-TTS maintains speech quality. However, further increases in the compression threshold yield diminishing returns in acceleration beyond T3. This highlights the intrinsic limitations of relying solely on the BS strategy for acceleration.


\textbf{BS strategy is essential for maintaining generation quality.} As shown in Figure~\ref{fig:ablation}, while only applying the TS strategy achieves greater acceleration of F5-TTS, it significantly degrades audio quality. For F5-TTS, employing TS alone leads to a substantial decrease in SIM-o at equivalent compression thresholds, with WER even reaching 23.06\% at the maximum threshold, indicating loss of conditional information during inference.
\textbf{The Check Phase and Pre-Calibration Phase are critical for identifying effective acceleration strategies.} As evidenced in Figure~\ref{fig:ablation}, omitting the first two phases of DiTReducio leads to significant degradation in both generation quality and inference speed for F5-TTS under equivalent compression thresholds. These findings confirm that these phases guide the Calibration Phase toward a superior strategy combination.

\section{Conclusion}
In this paper, we present a training-free acceleration approach for DiT-based TTS models, which derives a persistent acceleration strategy through a progressive calibration process. We observe temporal redundancy and branch redundancy in model inference, and develop corresponding strategies to exploit them effectively. We analyze the relationship between attention patterns and temporal redundancy in a specific layer-step pair, and develop a calibration framework that effectively identifies and compresses internal redundancies in a model-specific manner. Our experimental results verify that DiTReducio reduces computational costs in both attention and feed-forward modules while maintaining generation quality and compatibility with efficient attention computation libraries such as FlashAttention.

\vspace{4em}

\section*{Limitations}
Firstly, the applicability of DiTReducio is primarily constrained to DiT-based speech synthesis models, which limits its generalization to other model architectures.  Additionally, the framework demands high-quality calibration audio for optimal performance. 


\bibliography{custom}
\appendix
\clearpage
\section{Appendix}
\label{sec:appendix}

\subsection{Implementation Details}
\label{subsec:implementation}
For F5-TTS, we optimize the DiT implementation by integrating the conditional and unconditional branch computations into a single forward pass. Specifically, we concatenate conditional and unconditional inputs into a single batch, with the first half used for conditional inputs and the second half for unconditional ones. This modification allows the BS strategy to be applied while retaining functional equivalence with the original dual-pass method.

For MegaTTS 3, which employs multi-condition classifier-free guidance, we adapted the BS strategy to accommodate its dual conditional branches by computing two separate residuals: one between the text conditional branch and the speaker conditional branch, and another between the unconditional branch and the speaker conditional branch. During inference, only the speaker conditional branch is computed explicitly, while the other two branches are reconstructed by adding their respective residuals to it.

Both models are enhanced with FlashAttention, confirming the compatibility of our framework with efficient attention implementations.

\subsection{Redundancy in Model}
\label{sec:app_red}
\begin{center}
    \begin{figure*}[t]
    \begin{subfigure}[t]{0.5\linewidth}
      \includegraphics[width=0.9\columnwidth]{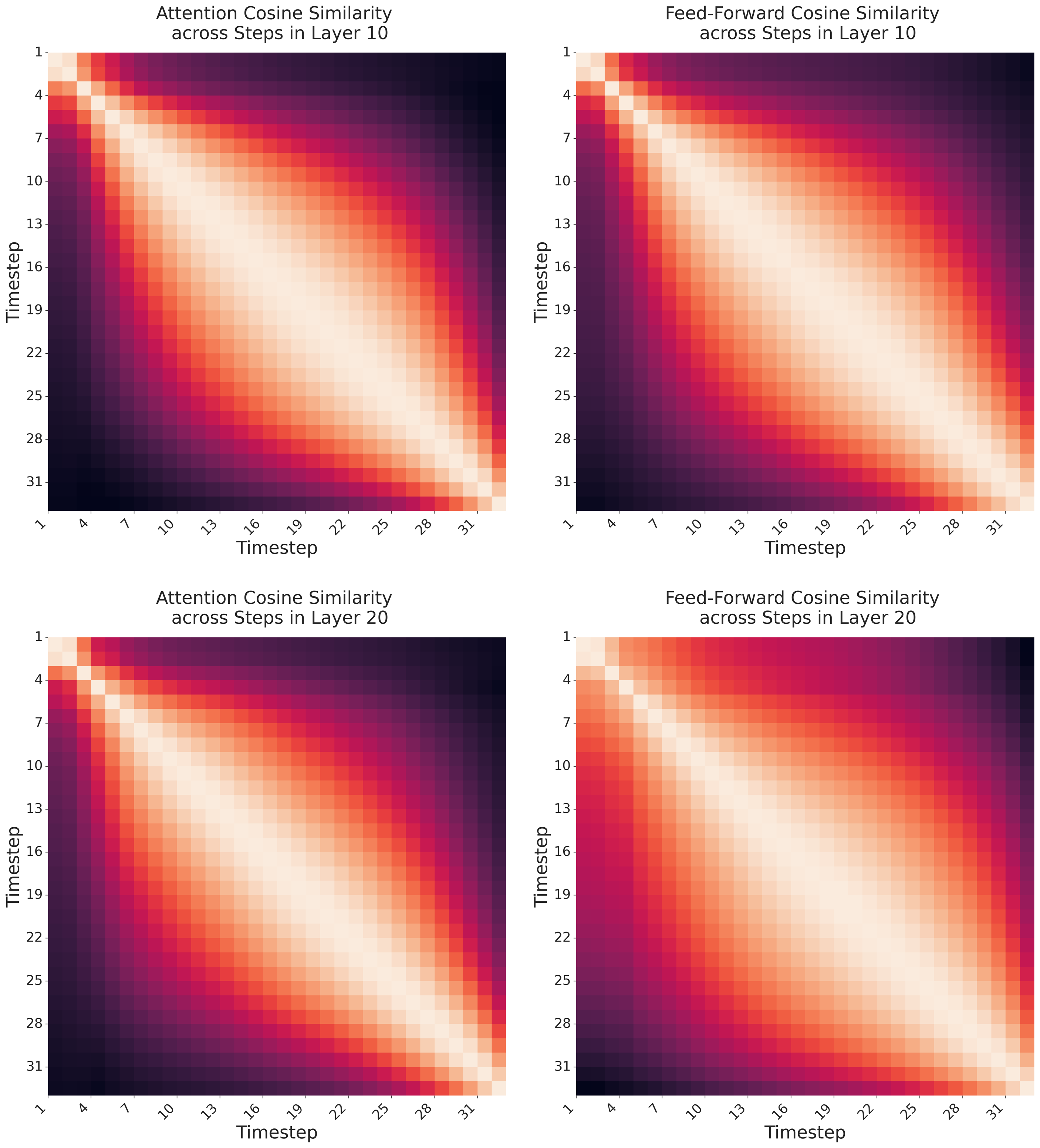}
      \caption{F5-TTS}
    \end{subfigure}
    \begin{subfigure}[t]{0.5\linewidth}
      \includegraphics[width=0.9\columnwidth]{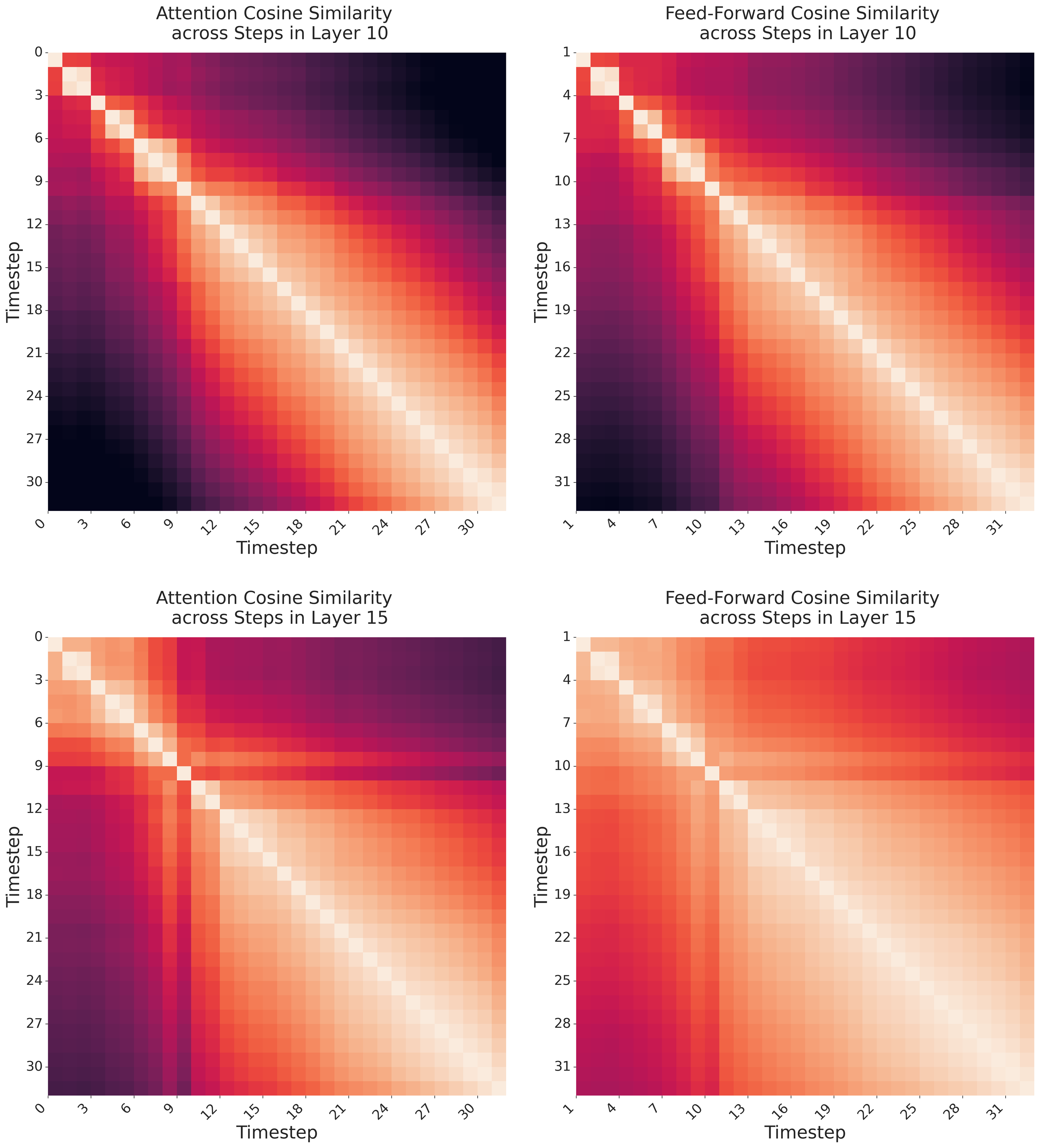}
      \caption{MegaTTS 3}
    \end{subfigure}
  \caption {\textbf{Temporal redundancy in F5‑TTS and MegaTTS 3.}}
  \label{fig:tr-of-models}
\end{figure*}
\end{center}

We analyze the temporal and branch redundancy in both F5-TTS and MegaTTS 3 models. Figure~\ref{fig:tr-of-models} presents the temporal redundancy analysis results. For F5-TTS, we examine layers 10 and 20, computing the cosine similarity between outputs from adjacent denoising timesteps for both attention and feed-forward. Similar analysis is conducted for MegaTTS 3, focusing on layers 10 and 22. The results reveal high temporal redundancy during the inference of the model.

\begin{center}
\begin{figure*}[h]
    \begin{subfigure}[t]{0.5\linewidth}
        \centering
        \includegraphics[width=0.47\columnwidth]{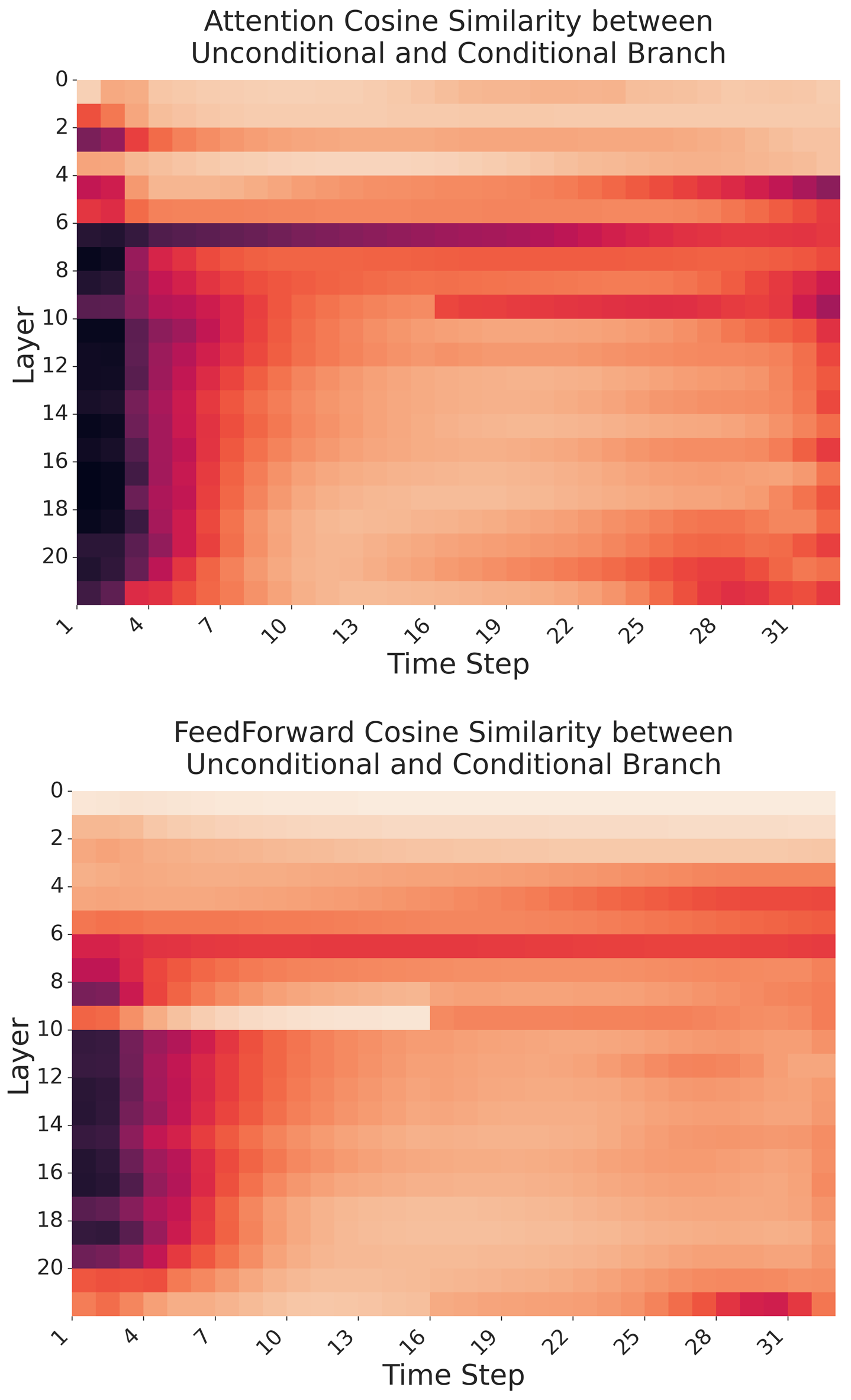}
        \caption{F5-TTS}
    \end{subfigure}
    \begin{subfigure}[t]{0.5\linewidth}
        \centering
        \includegraphics[width=0.9\columnwidth]{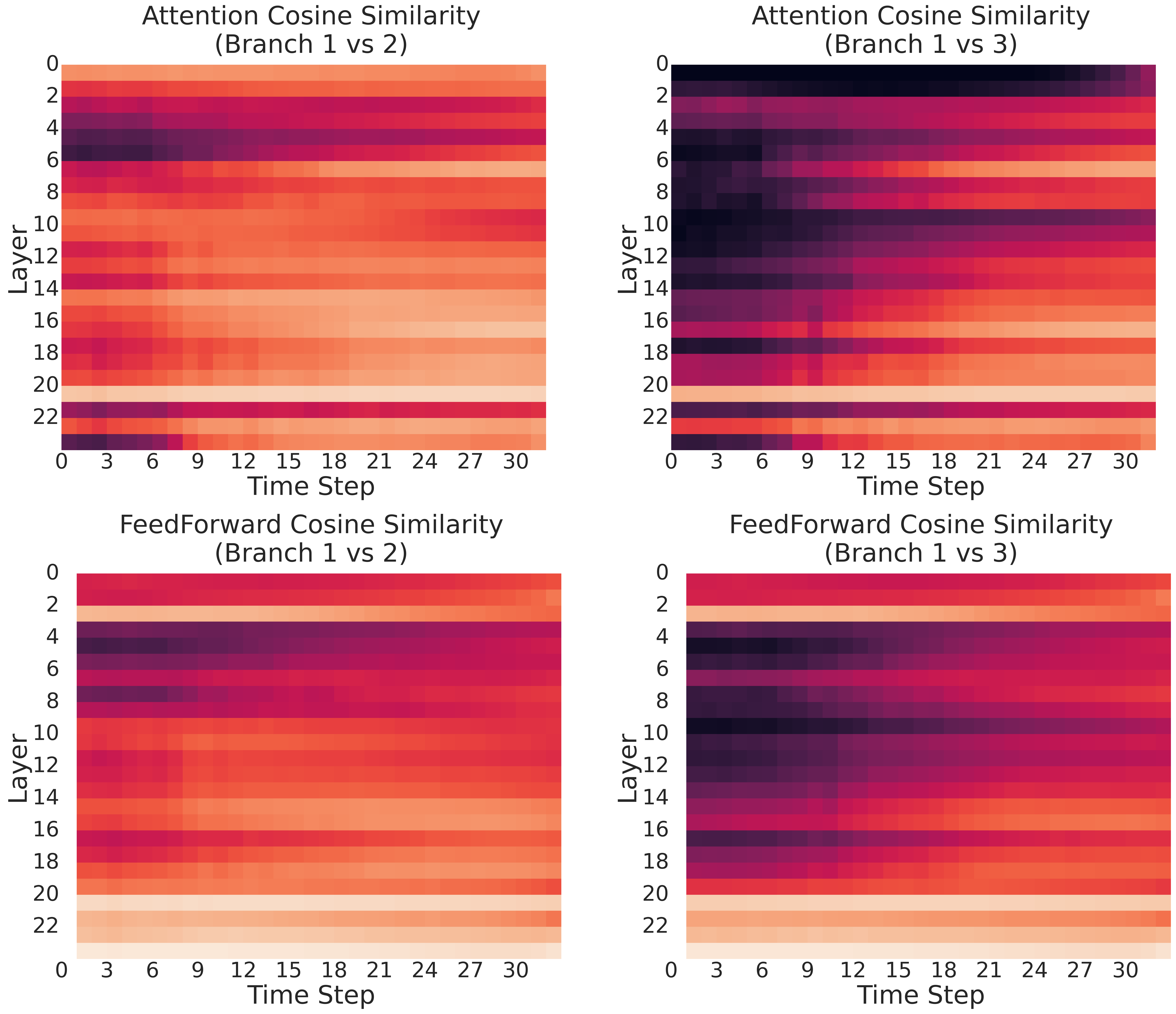}
        \caption{MegaTTS 3}
    \end{subfigure}
    \caption{\textbf{Branch redundancy in F5‑TTS and MegaTTS 3.}}
    \label{fig:br-of-models}
\end{figure*}
\end{center}

The branch redundancy characteristics are depicted in Figure~\ref{fig:br-of-models}. For F5-TTS, we measure the cosine similarity between outputs from conditional and unconditional branches for both attention and feed-forward modules across various timesteps. For MegaTTS 3, which utilizes multi-condition classifier-free guidance with two conditional branch outputs, we analyze the similarities among all three branches, specifically comparing the speaker conditional branch (Branch 1) with both the text conditional branch (Branch 2) and the unconditional branch (Branch 3).

\subsection{Method Distribution}

\begin{figure*}[!htbp]
  \centering
  \includegraphics[width=\textwidth]{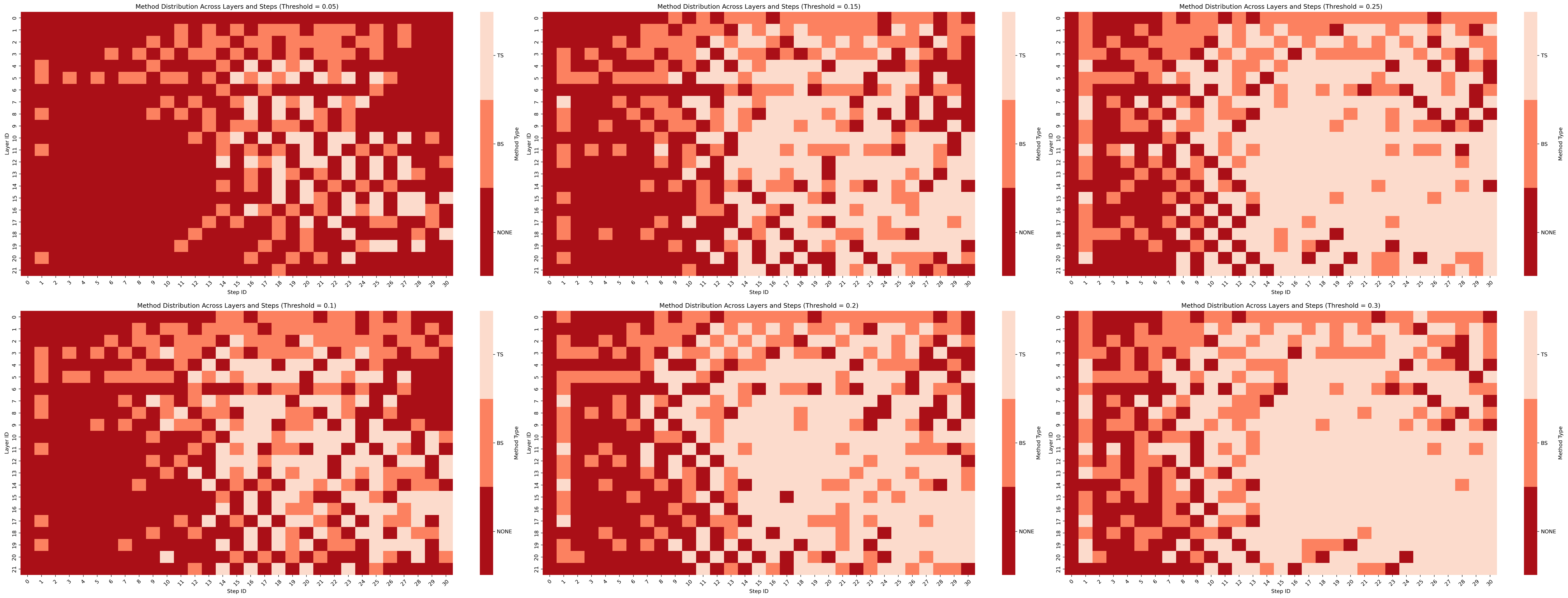}
  \caption {\textbf{Method Distribution of F5-TTS across compression thresholds}.}
  \label{fig:appendix_method-dist}
\end{figure*}

Figure~\ref{fig:appendix_method-dist} shows the method distribution heatmaps from the calibration of F5-TTS under increasing compression thresholds. Each cell in the heatmaps represents the compression method applied to a specific layer-step pair. The heatmaps are arranged in a grid layout from left to right and top to bottom, with each subplot corresponding to a threshold in 0.05 increments, ranging from 0.05 to 0.30. As the compression threshold rises, we observe a progressive increase in the number of layer-step pairs employing compression strategies. Notably, layer-step pairs at later timesteps are more likely to adopt TS, whereas those at earlier timesteps tend to remain uncompressed or use BS. This suggests an underlying connection between internal model redundancy and the progression across diffusion timesteps.

\subsection{Potential Risks}

While DiTReducio offers an efficient, training-free approach to accelerate DiT-based TTS models, its deployment entails certain risks that require careful management. A key concern is the potential degradation of speech quality in high-stakes applications such as healthcare, legal transcription, or emergency response. Even minor distortions could lead to misinterpretation or diminished user trust, highlighting the necessity of thorough validation in critical domains. Additionally, the enhanced efficiency of DiTReducio may lower the barrier for misuse, enabling malicious actors to generate deceptive audio content such as deepfakes or impersonation attacks. Robust safeguards are therefore essential to mitigate such risks.

\end{document}